\documentclass[10pt,conference]{IEEEtran}
\usepackage{amsmath}
\usepackage{amssymb}
\usepackage{amsfonts}
\usepackage{bbm}

\newcommand{\Rmnum}[1]{\expandafter\@slowromancap\romannumeral #1@}
\usepackage{graphicx}
\usepackage{epsfig}
\usepackage{subfigure}
\usepackage[sort]{cite}
\usepackage{xcolor}
\usepackage{enumerate}
\usepackage{extarrows}
\usepackage{algorithmic}
\usepackage{subfigure}
\usepackage[lined,ruled,linesnumbered]{algorithm2e}
\usepackage{tensor}
\usepackage{balance}
\usepackage{graphicx}
\usepackage[draft,bookmarks=false]{hyperref}
\def\BibTeX{{\rm B\kern-.05em{\sc i\kern-.025em b}\kern-.08em
		T\kern-.1667em\lower.7ex\hbox{E}\kern-.125emX}}

\begin{document}

\title{ Reliable Near-Field Multi-User Positioning Informed by Two-Stage MUSIC}

\author{\IEEEauthorblockN{ Jiaying Li${}^{\ast}$, Haifeng Wen${}^{\ast}$, Changsheng You${}^{\text { § }}$, Yuanwei Liu${}^{\dagger}$, and Hong Xing${}^{\ast\ddagger}$
}\\
	\IEEEauthorblockA{${}^\ast$ The Hong Kong University of Science and Technology (Guangzhou), Guangzhou, China \\
    $\text { § }$ Southern University of Science and Technology, Shenzhen, China \\
    ${}^\dagger$ The University of Hong Kong, HK SAR, China\\
    ${}^\ddagger$ The Hong Kong University of Science and Technology, HK SAR, China \\
		E-mails:~\{jli989, hwen904\}@connect.hkust-gz.edu.cn, youcs@sustech.edu.cn,~yuanwei@hku.hk,~hongxing@ust.hk
  }
  	\thanks{The work of H. Xing was supported in part by the Guangdong Basic and Applied Basic Research Foundation under Grant 2025A1515010123, and in part by the Guangdong Provincial Key Lab of Integrated Communication, Sensing and Computation for Ubiquitous Internet of Things under Grant 2023B1212010007.}}

\maketitle
\thispagestyle{empty}
\IEEEpeerreviewmaketitle

\newtheorem{definition}{\underline{Definition}}[section]
\newtheorem{fact}{Fact}
\newtheorem{assumption}{Assumption}
\newtheorem{theorem}{\underline{Theorem}}[section]
\newtheorem{lemma}{\underline{Lemma}}[section]
\newtheorem{proposition}{\underline{Proposition}}[section]
\newtheorem{corollary}[proposition]{\underline{Corollary}}
\newtheorem{example}{\underline{Example}}[section]
\newtheorem{remark}{\underline{Remark}}[section]

\newcommand{\mv}[1]{\mbox{\boldmath{$ #1 $}}}
\newcommand{\mb}[1]{\mathbb{#1}}
\newcommand{\Myfrac}[2]{\ensuremath{#1\mathord{\left/\right.\kern-\nulldelimiterspace}#2}}
\newcommand\Perms[2]{\tensor[^{#2}]P{_{#1}}}

\begin{abstract}
Near-field localization is a promising technique for high-resolution multi-user positioning in future wireless systems, but its performance is often degraded by scattering-induced coherent propagation. Existing near-field localization methods, which require separate parameter estimation and path/source association, suffer from high computation overhead and accumulated errors, and usually do not provide any guarantee on reliability. In this paper, we propose \emph{MUSIC-Net}, an end-to-end near-field positioning deep learning (DL) framework informed by two-stage MUltiple SIgnal Classification (MUSIC) in mixed line-of-sight (LoS) and non-LoS (NLoS) multi-path scenarios, which embeds the two-stage MUSIC objects into training to isolate the LoS-related signal subspace and to identify a surrogate distance. The proposed framework directly recovers multi-user positions without the need for involved NLoS parameter estimation or path/source association. Furthermore, we introduce split conformal prediction (SCP) to move beyond point-estimation-based positioning towards statistically guaranteed (confidence) set estimation for all users. Numerical results show that the proposed MUSIC-Net achieves lower mean positioning error (MPER) than existing benchmarks and yields tighter SCP-calibrated prediction regions, demonstrating both accurate LoS localization and efficient uncertainty quantification (UQ) in coherent multi-path environments.
\end{abstract}


\section{Introduction}
\label{sec:introduction}

With rapid development of extremely large-scale antenna arrays and higher-band radio-frequency (RF) communications, next-generation wireless systems are envisioned to operate in the radiative near-field regime featuring spherical wavefronts~\cite{liu2023near}, which entails involved multi-path channel estimation. Meanwhile, unlike conventional far-field localization, which mainly resolves angular information under the planar-wave assumption, near-field localization benefits from such high-resolution estimation of both direction of arrival (DoA)  and distance of source targets, thus enabling high-precision near-field multi-user positioning \cite{lei2025near_field}.

Existing near-field localization methods mainly comprise beamforming~\cite{yang2025beamforming}, compressive sensing~\cite{rinchi2022cs}, subspace-based~\cite{huang2002near,liang2010twostage,zhang2018rdmusic,qu2024two}, and learning-based approaches~\cite{su2021cnn_near,li2025unsupervised,Ji2024trans_near,Gast2026NFsubspace,gast2025dcd}. Among them, subspace-based methods such as 2D MUltiple SIgnal Classification (MUSIC)  and its variants are particularly appealing, as they provide both high resolution and physical interpretability~\cite{huang2002near,liang2010twostage,zhang2018rdmusic,qu2024two}. In particular, 2D MUSIC~\cite{huang2002near} performed near-field localization through a two-dimension angle-range spectral search, whereas its multi-stage variants, including two-step MUSIC~\cite{liang2010twostage}, RD-MUSIC~\cite{zhang2018rdmusic}, and two-stage MUSIC~\cite{qu2024two}, reduced the computational overhead by decomposing the original joint angle-range search into multiple stages of one-dimension search. However, two-step MUSIC~\cite{liang2010twostage} and RD-MUSIC~\cite{zhang2018rdmusic} require restrictive antenna spacing conditions, while two-stage MUSIC~\cite{qu2024two} avoids imposing this limitation by introducing
a surrogate distance  for angular search, which decomposed  the original
two-dimension angle-distance search into two sequential
one-dimension searches.

On another front, recent deep learning (DL) based methods improve positioning robustness in challenging propagation conditions such as low signal-to-noise ratio (SNR) regimes, model mismatch, and multi-path environments~\cite{su2021cnn_near,li2025unsupervised,Ji2024trans_near}. Among these prior arts,  end-to-end (E2E) approaches aim for directly learning the mapping from measurements  to  positions using, e.g., convolutional neural networks (CNNs) \cite{su2021cnn_near,li2025unsupervised} and Transformers \cite{Ji2024trans_near}, but often rely on highly parameterized DL models with limited generalization capability. Meanwhile, the model-based DL methods \cite{Gast2026NFsubspace,gast2025dcd} design classical signal processing algorithms informed neural-network architectures, thereby combining data-driven flexibility with physical interpretability.  {Prototypical} examples include NF-SubspaceNet~\cite{Gast2026NFsubspace}, which adopted a surrogate covariance matrix for subsequent 2D MUSIC processing, and DCD-MUSIC~\cite{gast2025dcd}, which learned separate surrogate covariance matrices for angle and distance estimation, respectively.

However, all the above methods may suffer from scattering-induced multi-paths, of which the line-of-sight (LoS) path and the non-line-of-sight (NLoS) paths associated with the same user can be highly coherent, thus distorting the structure of the LoS-only induced subspace and making conventional localization methods demanding if not impossible~\cite{wen2019survey}. 
Existing methods for coherent-source localization mainly start with estimating all path-related parameters, and then performing further association schemes to group those multi-paths associated with the same source in a multi-user setting \cite{xie2017source}. 
However, subsequent LoS-path identification for each group entails additional computation overhead and possibly accumulated positioning error. 
These limitations motivate a near-field multi-user positioning scheme that directly targets at LoS parameters estimation bypassing the coherent NLoS estimation and subsequent path/source association steps.

Furthermore, the reliability of the positioning scheme also plays a crucial role especially in safety-critical localization, such as autonomous driving and tactile networks.
Split conformal prediction (SCP)~\cite{angelopoulos2023conformal} is known to provide a model-agnostic, distribution-free and training-free uncertainty quantification (UQ) framework that moves beyond point-parameter estimation by constructing {\it post-hoc} confidence sets with statistically guaranteed true-position coverage. For instance, SCP has recently found its applications in, e.g., channel prediction~\cite{cohen2023calibrating}, wireless resource allocation~\cite{binucci2025conformal}, 
and beam selection~\cite{deng2026scan}. Based on the motivations described above, the contributions of this paper are summarized as follows:
\begin{itemize}
    \item We propose \emph{MUSIC-Net}, a near-field multi-user positioning framework in coherent multi-path conditions, which introduces two learning modules to estimate the LoS-related subspace and the surrogate distance, respectively,  while preserving the interpretability of the two-stage MUSIC, enables users' positioning without explicit NLoS parameter estimation or path/source association.
\item We integrate SCP into the proposed multi-user positioning framework to construct a robust position set against uncertainties underlying MUSIC-Net inference with statistical coverage guarantees for all users.
    \item Numerical results show that MUSIC-Net achieves the lowest mean positioning error (MPER) across different SNR regimes and yields the tightest prediction regions among the benchmarks, demonstrating reliable and efficient UQ in coherent multi-path environments.
\end{itemize}


\section{System Model and Reliable Problem Formulation}
\label{sec:system_model}
\subsection{System Model}
\label{subsec:system_model}
\begin{figure}[t]
\centering
\includegraphics[width=3in]{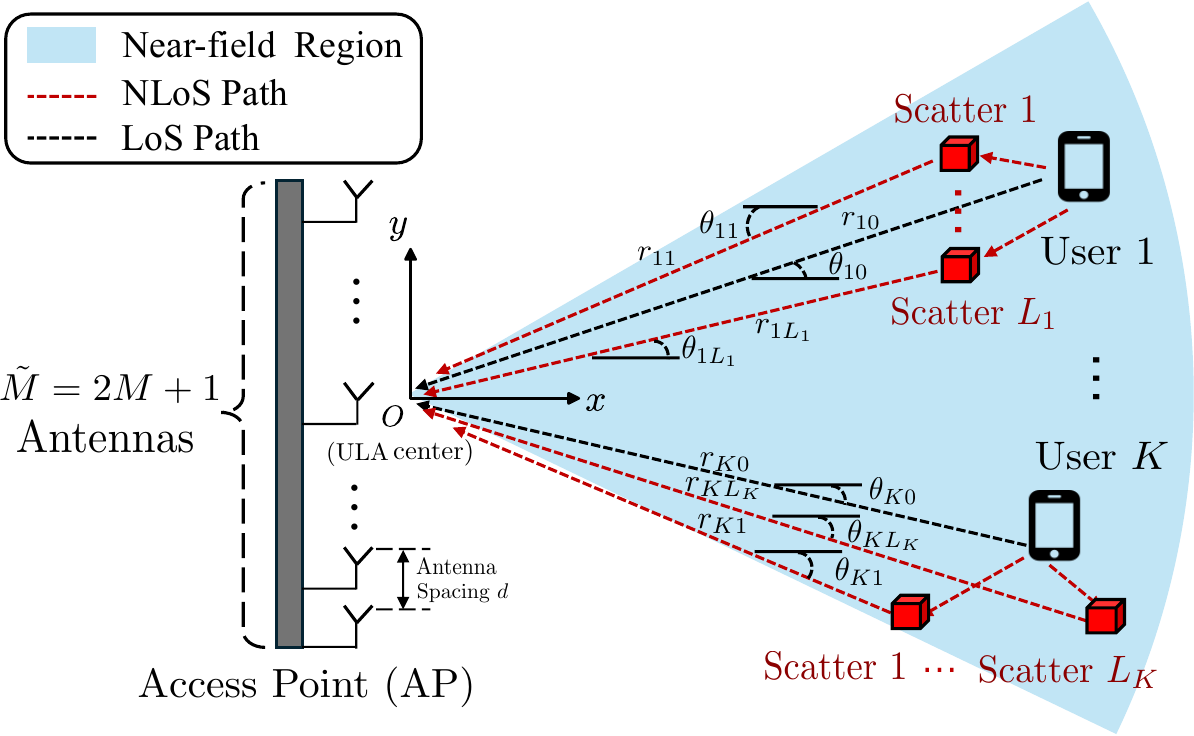}
\caption{An illustration of the system for near-field multi-user positioning with multiple coherent NLoS paths.}
\vspace{-0.2in}
\label{fig:system_model}
\end{figure}

As shown in Fig.~\ref{fig:system_model}, we consider a multi-user narrowband near-field {channel} for uplink localization, where an access point (AP) equipped with a uniform linear array (ULA) with $\tilde{M}=2M+1$ antennas receives signals from $K$ single-antenna user equipments (UEs), denoted by $k\in\{ 1,...,K\}\triangleq [K]$. The users are assumed to lie in the near-field region, i.e., the distance $r$ from users to the center of the ULA  satisfying $r<2D^2/\lambda$, where $D=(\tilde{M}-1)d$ is the array aperture; $d$ is the antenna spacing; and $\lambda$ is the carrier wavelength. 

\label{sec:system_model_problem_formulation}

Under the narrowband near-field assumptions, the steering vector $\boldsymbol a(\theta,r)\in\mathbb{C}^{\tilde{M}\times 1}$ for a target user at DoA $\theta$ and distance $r$ is given by \cite{liu2023near}
\begin{equation}
\boldsymbol a(\theta,r)=
\left[
e^{-j\frac{2\pi}{\lambda}\|\boldsymbol p-\boldsymbol s_{-M}\|_2},
\ldots,
e^{-j\frac{2\pi}{\lambda}\|\boldsymbol p-\boldsymbol s_{M}\|_2}
\right]^T,
\label{eq:steering_vector}
\end{equation}
where $\boldsymbol p=r[\cos\theta,\, \sin\theta]^T$ and $\boldsymbol s_n=[0,nd]^T$ denote the positions of the user and the $n$-th antenna element, respectively.
We assume that, for the signal transmitted by the $k$-th user, there exists one LoS propagation path and $L_k\ge 0$ {NLoS ones} generated by scatterers. Let $s_k(t)$ denote the signal transmitted by user $k\in[K]$ at time slot $t\in \{1,\ldots,T\} \triangleq [T]$, where $T$ is the total number of time slots with $T>K$. We also  assume that {$s_k(t)$ is independent, denoted by $s_k(t) \sim \mathcal{CN}(0, P_k)$}. Then, the received signal $\mv y(t) \in \mathbb{C}^{\tilde M \times 1}$ can be expressed as
\begin{equation} 
\boldsymbol y(t)=\boldsymbol y_{\mathrm{LoS}}(t)+\boldsymbol y_{\mathrm{NLoS}}(t)+\boldsymbol n(t) \label{eq:received_signal} 
\end{equation}
where 
\begin{equation}
\boldsymbol y_{\mathrm{LoS}}(t)= \sum_{k=1}^{K}\boldsymbol a(\theta_{k0},r_{k0})s_k(t),
\label{eq:los_component}
\end{equation}
and 
\begin{equation}
\boldsymbol y_{\mathrm{NLoS}}(t)= \sum_{k=1}^{K}\sum_{l=1}^{L_k}\boldsymbol a(\theta_{kl},r_{kl})c_{kl}s_k(t),
\label{eq:nlos_component}
\end{equation}
respectively, {where $\theta_{k0}$ and $r_{k0}$ denote the DoA and  distance associated with the LoS path of the $k$-th user}; $\theta_{kl}$, $r_{kl}$, and $c_{kl}$ denote the DoA, distance, and complex reflection coefficient associated with the  $l$-th NLoS path of the $k$-th user, where $l \in \{1,..,L_k \} \triangleq [L_k]$ and $k \in [K]$; and $\boldsymbol n(t)\sim \mathcal{CN}(\boldsymbol 0,\sigma^2\boldsymbol I)$ is the additive white Gaussian noise (AWGN). Accordingly, the covariance matrix of $\boldsymbol y(t)$ is given by
\begin{equation}
\boldsymbol R = \mathbb{E} [\boldsymbol y(t)\boldsymbol y^H(t)]
= \boldsymbol R_{\mathrm{LoS}}+\boldsymbol R_{\mathrm{NLoS}}+\boldsymbol R_{\mathrm{cross}}+\sigma^2\boldsymbol I,
\label{eq:all_covariance}
\end{equation}
where $\boldsymbol R_{\mathrm{LoS}} = \mathbb{E}[\boldsymbol y_{\mathrm{LoS}}(t)\boldsymbol y_{\mathrm{LoS}}^H(t)]$, $\boldsymbol R_{\mathrm{NLoS}} = \mathbb{E}[\boldsymbol y_{\mathrm{NLoS}}(t)\boldsymbol y_{\mathrm{NLoS}}^H(t)]$, and $\boldsymbol R_{\mathrm{cross}} = \mathbb{E}[\boldsymbol y_{\mathrm{LoS}}(t)\boldsymbol y_{\mathrm{NLoS}}^H(t)] + \mathbb{E}[\boldsymbol y_{\mathrm{NLoS}}(t)\boldsymbol y_{\mathrm{LoS}}^H(t)]$ denote the covariance matrices of the LoS and NLoS components and their cross-correlation term, respectively. In practice, one can approximate the true covariance matrix $\boldsymbol{ R}$  {by} the sample covariance matrix (SCM)  {as}
\begin{equation}
    \boldsymbol{\widehat R} = \frac{1}{T}{\boldsymbol Y} {\boldsymbol Y}^H  \label{eq:SCM} %
\end{equation}

where $\boldsymbol Y=[\boldsymbol y(1),\ldots,\boldsymbol y(T)] \in \mathbb{C}^{\tilde M \times T}$ is the observations matrix over $T$ consecutive time slots.

\subsection{Reliable  {Multi-User Positioning}}
\label{subsec:joint_estimation}
Based on the observation model in \eqref{eq:received_signal},  {we aim for estimating all $K$ users' }location-related parameters  {associated with the LoS paths, i.e.,  $\boldsymbol{\Theta}_{\mathrm{LoS}} = (\theta_{k0},r_{k0})_{k=1}^{K}$.}
We assume that the number of users $K$ is known {\it a priori},  {which can be estimated} by conventional methods such as Akaike information criterion (AIC) and minimum description length (MDL)~\cite{Gast2026NFsubspace}. 
The parameters associated with the NLoS paths, i.e., $\boldsymbol{\Theta}_{\mathrm{NLoS}} = ((\boldsymbol{\Phi}_k)_{k=1}^{K},\sigma^2)$ where $\boldsymbol{\Phi}_k= (\theta_{kl},r_{kl},c_{kl})_{l=1}^{L_k}$, are treated as nuisance parameters.
 {To estimate the location-related parameters $\mv\Theta_{LoS}$, the standard} approach is to apply subspace-based methods such as the 2D MUSIC algorithm \cite{huang2002near} and its variants \cite{qu2024two,zhang2018rdmusic} directly to the SCM $\boldsymbol {\widehat R}$, which generally work well in LoS only or LoS dominant scenarios. However, in the considered multi-path scenario, the term $\boldsymbol R_{\mathrm{NLoS}}+\boldsymbol R_{\mathrm{cross}}$ in \eqref{eq:all_covariance} deteriorates the subspace structure induced solely by the LoS components.   {Besides, the estimated locations $\widehat{\boldsymbol{p}}_k$ of all users \(k\in[K]\), where $\boldsymbol{\widehat p}_k = \widehat r_{k0}
[\cos(\widehat \theta_{k0}), \sin(\widehat\theta_{k0})]^T$, also deviate from their true positions $\boldsymbol{p}_k, k\in[K]$, due to random factors including, e.g., remained NLoS traces, AWGN, and the positioning algorithms due to the limited number of observations.}

{ {
In this paper, we develop in scenarios with multiple coherent NLoS paths a multi-user positioning scheme that is robust against uncertainty induced by channel modeling or localization methods.}
Specifically, 
 {we are interested in constructing a collection of confidence sets 
$\{\mathcal C(\boldsymbol{\widehat p}_k)\}_{k=1}^{K}$ such that they jointly cover the true positions of the $K$ users with probability at least $1-\alpha$, i.e.,}
\begin{equation}
\mathbb P\!\left(
\exists\,\pi\in\Pi_K:
\boldsymbol{p}_k \in 
\mathcal C(\boldsymbol{\widehat p}_{\pi(k)}),
~\forall k\in[K]
\right)\ge 1-\alpha,
\label{eq:marginal_joint_coverage}
\end{equation}
where $\Pi_K$ denotes the set of all permutations of $[K]$, and the probability is taken with respect to the joint distribution of the test data and calibration data, which will be elaborated in later sections. }

\section{Two-stage MUSIC Informed MUSIC-Net}
\label{sec:subs_NN}
To address the near-field multi-user positioning with multiple coherent NLoS paths, we propose \emph{MUSIC-Net}, a deep learning framework informed by the \emph{two-stage MUSIC} algorithm \cite{qu2024two}. 
\subsection{Motivation}
In standard two-stage MUSIC algorithm \cite{qu2024two}, since the LoS paths of the users dominate uplink transmissions, the covariance matrix of the received signal $\mv R$, estimated by the SCM $\widehat{\mv R}$, can be well approximated by \(\boldsymbol{R}=\sum_{k=1}^K P_k \boldsymbol{a}(\theta_{k0},r_{k0})  \boldsymbol{a}^H(\theta_{k0},r_{k0})  +\sigma^2\boldsymbol{I}\) in~\eqref{eq:all_covariance}, and therefore the eigenspace of $\mv R$ is decomposed, by eigenvalue decomposition (EVD), into a signal subspace $\mv U_s$ and a noise subspace $\mv U_n$ as assumed in conventional far-field MUSIC algorithm. In a near-field setting, the MUSIC based multi-user localization problem thus translates into a two-dimension search for the top-$K$ largest peaks based on the spectrum function given by
\begin{align}
P(\theta, r)= \frac{1}{\|\boldsymbol{a}^H \left( \theta,r\right) \boldsymbol{U}_n\|_2}. \label{eq:hehe} 
\end{align}

Instead of an exhaustive two-dimension search over $[-\tfrac{\pi}{2}, \tfrac{\pi}{2}]\times[r_{\min}, r_{\max}]$, two-stage MUSIC decouples the above search into two one-dimension spectral searches, and performs them in the following two stages. 1) \textbf{DoA search}: Based on a surrogate distance $r_0$, which is obtained by either  maximizing the minimum correlation (MMC) rule that maximizes the worst-case steering vector correlation or  exceeding the minimum correlation threshold (EMCT) rule that enforces this correlation to exceed a prescribed threshold \cite{qu2024two}, estimate the DoAs of the users by searching for the $K$ peaks with $r$ replaced by $r_0$ in \eqref{eq:hehe}; and 2) \textbf{Distance search}: Based on the estimated DoAs $\widehat{\theta}_{k0}$, $k\in[K]$, estimate the distance corresponding to each user $k\in[K]$ by searching for one peak with $\theta$ replaced by $\widehat{\theta}_{k0}$ in \eqref{eq:hehe}.

However, we cannot trivially tailor the two-stage MUSIC algorithm to multi-user localization with NLoS, as the structure of the $\mv R$ becomes fundamentally different from $\boldsymbol{R}_{\text{LoS}}$ due to involved unknown NLoS parameters (cf. \eqref{eq:all_covariance}), thus disabling direct estimation of the noise subspace $\mv U_n$, which motivates a novel data-driven design as introduced in the next subsection.



\subsection{MUSIC-Net Training}
\begin{figure}[thp]
\centering
\includegraphics[width=3.2in]{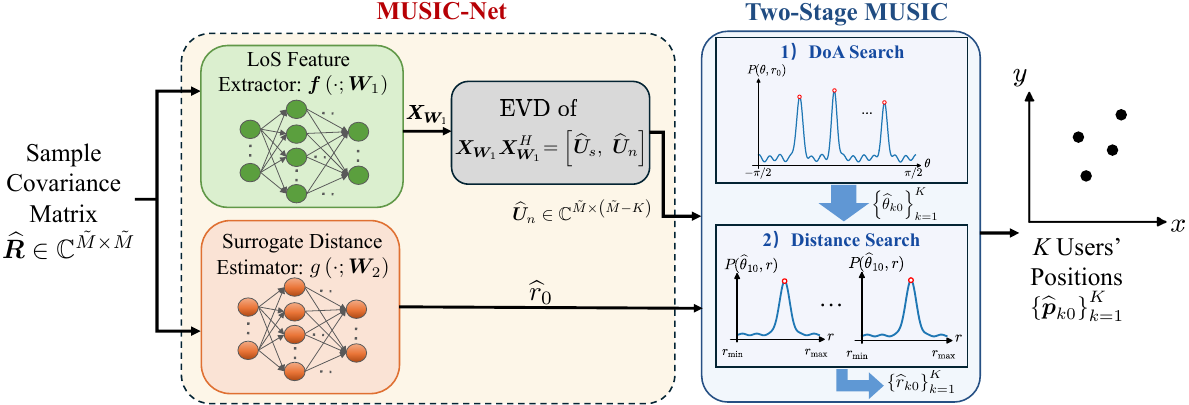}
\caption{ {The proposed two-stage MUSIC informed MUSIC-Net.}}
\label{fig:proposed_architecture}
\end{figure}

We propose a novel multi-user localization scheme enabled by {MUSIC-Net}, which substitutes two dedicated trainable modules for the estimated noise subspace $\widehat{\mv U}_n$ and the surrogate distance $\hat{r}_0$ in the standard two-stage MUSIC, respectively. As illustrated in Fig. \ref{fig:proposed_architecture}, prior to performing the two-stage MUSIC procedure, we adopt MUSIC-Net training that consists of 1) a LoS feature extractor module and 2) a surrogate distance estimation module to obtain an LoS-plus-noise subspace estimation $\widehat{\mv U}_{\!n}$ and a \emph{surrogate distance} estimation $\widehat{r}_0$, respectively, from the input SCM $\widehat{\mv R}$.

Specifically, the LoS feature extractor,  denoted by $\boldsymbol f(\cdot;\boldsymbol W_{\!1})$ with trainable parameters $\boldsymbol W_{\!1}$, maps  the input $\widehat{\boldsymbol R}$ to a complex-valued feature matrix $\boldsymbol X_{\!\boldsymbol W_{\!1}}\in\mathbb C^{\tilde M\times \tilde M}$, i.e., $\boldsymbol f(\widehat{\boldsymbol R};\boldsymbol W_{\!1})=\boldsymbol X_{\!\boldsymbol W_{\!1}}$, which predicts denoised signals induced by a virtual subspace dominated by the LoS paths. Then, we perform the EVD on $\boldsymbol X_{\!\boldsymbol W_{\!1}}\boldsymbol X_{\!\boldsymbol W_{\!1}}^H$:
\begin{equation}
\boldsymbol X_{\!\boldsymbol W_{\!1}} \boldsymbol X_{\!\boldsymbol W_{\!1}}^H
=
\left[
\widehat{\boldsymbol U}_{\! s},~
\widehat{\boldsymbol U}_{\!n}
\right]
\widehat{\boldsymbol \Sigma}
\left[
\widehat{\boldsymbol U}_{\!s},~
\widehat{\boldsymbol U}_{\!n}
\right]^H,
\label{eq:evd_of_surrogate}
\end{equation}
where $\widehat{\boldsymbol U}_{\! s}\in\mathbb C^{\tilde M\times K}$ contains the eigenvectors associated with the $K$ largest eigenvalues spanning over the estimated LoS-induced subspace; $\widehat{\boldsymbol U}_{\!n}\in\mathbb C^{\tilde M\times(\tilde M-K)}$ contains the remaining eigenvectors; and $\widehat{\boldsymbol \Sigma}$ is a diagonal matrix containing all eigenvalues in descending order. 
Next, we adopt the subspace distance \cite{chen_2025_srl} between the target LoS-induced signal subspace $\mv U_{\!s}$, which is obtained by EVD on $\boldsymbol A_{\mathrm{LoS}}\boldsymbol A_{\mathrm{LoS}}^H$ with $\boldsymbol A_{\mathrm{LoS}}$ composed of steering vectors associated with only  LoS paths of all $K$ users, and the learned LoS-induced signal subspace $\widehat{\mv U}_{\!s}$ as the supervised (empirical) loss  given by
\begin{align}
\begin{aligned}
    \mathcal L_1 \left(\boldsymbol W_{\!1} \right)
=
\frac{1}{|\mathcal D|}
\sum_{\mathcal D}
\sum_{k=1}^{K}
\arccos\!\left(
\sigma_k\!\left(\boldsymbol U_s^H\widehat{\boldsymbol U}_{\! s}\right)
\right),
\end{aligned}
\label{eq:denoising}
\end{align}
where $\mathcal D$ is the training dataset composed of  $|\mathcal{D}|$ labelled data $(\widehat{\boldsymbol R},\boldsymbol\Theta_{\mathrm{LoS}})$, and $\sigma_k(\boldsymbol A)$ denotes the $k$-th largest singular value of the matrix $\boldsymbol A$. 
It is worth noting that there are other losses like SCM-fitting \cite{chen_2025_srl} and affine invariant distance \cite{barthelme2021doa}, which are, however, sensitive to SNR of the received signal, causing possibly bad generalization.


The surrogate distance estimator, denoted by $g(\cdot;\boldsymbol W_{\!2})$ with trainable parameters $\boldsymbol W_{\!2}$, maps the SCM $\widehat{\boldsymbol R}$ to a surrogate distance $\widehat{r}_0$, i.e., $ \widehat{r}_0=g(\widehat{\boldsymbol R};\boldsymbol W_{\!2}).$ 
To ensure that the surrogate LoS-induced subspace $\tilde{\mv U}_{\!s}$ associated with a single distance $\hat r_0$, which is obtained by EVD on $\mv A_{r_0}\mv A_{r_0}^H$ with $\boldsymbol A_{r_0} =
\big[ \boldsymbol a(\theta_{10},\widehat{r}_0),\ldots, \boldsymbol a(\theta_{K0},\widehat{r}_0) \big]$, aligns well with the estimated subspace $\widehat{\boldsymbol U}_{\!s}$ associated with the users' $K$ distances, the empirical training loss is therefore chosen as  the subspace distance between $\widetilde{\boldsymbol U}_{\!s}$ and $\widehat{\boldsymbol U}_{\!s}$, which is given by
\begin{align}
\begin{aligned}
     \mathcal L_2  \left(\boldsymbol W_{\!1},\boldsymbol W_{\!2}\right)
=
\frac{1}{|\mathcal D|}
\sum_{\mathcal D}
\sum_{k=1}^{K}
\arccos\left(
\sigma_k\!\left( \widetilde{\boldsymbol U}_{\!s}^H \widehat{\boldsymbol U}_{\!s}\right)
\right).
\end{aligned}
\label{eq:virtual_distance_learning}
\end{align}

The model parameters $\boldsymbol W_{\!1}$ and $\boldsymbol W_{\!2}$ of the two modules  are trained jointly, by minimizing the following combined multiple objectives:
\begin{equation}
\min_{\boldsymbol W_{\!1},\boldsymbol W_{\!2}}\;
\mathcal L_1(\boldsymbol W_{\!1})+\lambda\,\mathcal L_2(\boldsymbol W_{\!1},\boldsymbol W_{\!2}),
\label{eq:joint_train}
\end{equation}
where $\lambda>0$ is a hyperparameter that balances the loss induced by noisy NLoS paths and by surrogate distance estimation. 

In the inference stage, the MUSIC-Net takes an SCM sample $\widehat{\mv R}$ as an input and outputs the estimated $\widehat{\mv U}_{\!n}$ and $\widehat{r}_0 $, which are then fed into the two-stage MUSIC to obtain the estimates of DoA $\{ \widehat{\theta}_{k0}\}_{k=1}^K$ and distance $\{ \widehat{r}_{k0}\}_{k=1}^K$, resulting in estimated positions $\widehat{\boldsymbol{p}}_{k0} = \widehat r_{k0} [\cos(\widehat\theta_{k0}),\,\sin(\widehat\theta_{k0})]^T$  for all users in $[K]$.
However, like most DL-based localization methods, the proposed localization scheme produces only point estimates of users' positions without any robustness guarantee against uncertain behavior of MUSIC-Net due to limitations of training. In the next section, we augment point estimates with confidence sets.

\section{SCP based Positioning Calibration}
\subsection{Preliminary}
\label{subsec:SCP}
Among the various conformal prediction methods, SCP \cite{angelopoulos2023conformal} features simple implementation  and computational efficiency, thus being suitable for post-hoc calibration \cite{angelopoulos2023conformal}.  Suppose that we have a calibration dataset $\mathcal{D}_{\text {cal }}= \{(X^{(i)}, Y^{(i)})\}_{i=1}^n$, where $X^{(i)} \in \mathcal{X}$ and $Y^{(i)} \in \mathcal{Y}$, drawn independently and identically distributed (i.i.d.) from a distribution, denote the feature and the corresponding label, respectively. {Next}, given a prescribed \emph{nonconformity score} function $s: \mathcal{X} \times \mathcal{Y} \rightarrow \mathbb{R}$ associated with a predictive model, define the calibration score for each  data point  $(X^{(i)}, Y^{(i)})$ by $S^{(i)}=s\left(X^{(i)}, Y^{(i)}\right)$, which measures how poorly the  $Y^{(i)}$ conforms to the  $X^{(i)}$. {Then}, for a target mis-coverage level $\alpha \in(0,1)$ and a test feature $X^{(n+1)}$, SCP constructs the following prediction (confidence) set:
\begin{equation}
\mathcal C(X^{(n+1)})
= \{y \in\mathcal{Y} :s(X^{(n+1)},y)\le \widehat q_{1-\alpha}\},
\label{eq:CP_set_generation}
\end{equation}
which contains labels whose scores do not exceed the empirical quantile $\widehat q_{1-\alpha}$ given by
\begin{equation}
\widehat q_{1-\alpha}
=
Q\!\left(
(1-\alpha)\left(1+\Myfrac{1}{n}\right);
\{S^{(i)}\}_{i=1}^{n}
\right),
\label{eq:quantile_compute}
\end{equation}
where \(Q(\gamma;\{S^{(i)}\}_{i=1}^{n})\) is the \(\lceil n\gamma \rceil\)-th smallest value in the set  \(\{S^{(i)}\}_{i=1}^{n}\).

Under the assumption of exchangeable $\mathcal{D}_{\text{cal}}$ and $(X^{(n+1)}, Y^{(n+1)})$, the marginal coverage  of the SCP set \eqref{eq:CP_set_generation} is proven to satisfy $\mathbb P\!\left( Y^{(n+1)} \in \mathcal C({X}^{(n+1)} )\right)\ge 1-\alpha$~\cite{angelopoulos2023conformal}.

\subsection{Uncertainty Quantification for MUSIC-Net}
{
We provide in this subsection a UQ method for users' positions leveraging SCP. First, define  a held-out calibration dataset  {composed of $n$ samples}, denoted by
$\mathcal{D}_{\text{cal}}=\{\widehat{\boldsymbol{R}}^{(i)},  
\{(\theta_{k0}^{(i)},r_{k0}^{(i)})\}_{k=1}^{K^{(i)}}\}_{i=1}^n$, 
where $(\theta_{k0}^{(i)}, r_{k0}^{(i)})$ is the  {ground-truth DoA and distance associated with the LoS path of user  $k\in[K^{(i)}]$ in the $i$-th sample}. 

Next, we define the nonconformity score function $s(\cdot,\cdot)$ as  {
\begin{align}
s\left(
\widehat{\mv R},
\{(\theta_{k0},r_{k0})\}_{k=1}^{K}
\right)
=
\max_{k\in[K]}
\|
\mv p_{\pi^\star(k)}
-
\widehat{\mv p}_k
\|_2 ,
\label{eq:PER_score}
\end{align}}where $\widehat{\mv p}_{k}$ is the estimated position obtained by the proposed MUSIC-Net based positioning in Section~\ref{sec:subs_NN}, and $\pi^\star\in\Pi_K$ denotes the optimal permutation obtained by  {matching the estimated and ground-truth positions of the $K$ users leveraging} the permutation-based matching method, e.g., the Hungarian algorithm~\cite{kuhn1955hungarian}.}

{
 {Then}, for a  {newly} observed $K$-user SCM $\widehat{\mv R}$, we construct the  {confidence region} for each estimated user position based on the empirical quantile $\widehat{q}_{1-\alpha}$ obtained via \eqref{eq:quantile_compute} as
\begin{equation}
\mathcal C(\widehat {\mv p}_k)=
\left\{
\boldsymbol{{p}} \in \mathbb{R}^2
\mid
\left\| 
\boldsymbol{{p}} - \boldsymbol{\widehat{p}}_k
\right\|_2 
\le \widehat q_{1-\alpha}
\right\},
\quad k\in[K],
\label{eq:CP_set_construction}
\end{equation}
which corresponds to a circular  region centered on $\widehat{\boldsymbol p}_k$ with radius $\widehat q_{1-\alpha}$.
{ {Finally}, we  derive the following joint coverage guarantee for the
 {predicted} regions
$\{\mathcal C(\widehat{\mv p}_k)\}_{k=1}^{K}$.}


{
\begin{proposition}[ {Joint coverage guarantee}]
\label{prop:joint_coverage}
Assume  that the calibration dataset $\mathcal{D}_{\text{cal}}$ and an upcoming test sample $(\widehat{\boldsymbol{R}},  
\{(\theta_{k0},r_{k0})\}_{k=1}^{K})$ are exchangeable.
Then, the predicted regions
$\{\mathcal C(\widehat{\boldsymbol p}_k)\}_{k=1}^{K}$
constructed in \eqref{eq:CP_set_construction} satisfy  \eqref{eq:marginal_joint_coverage}.
\end{proposition}
\begin{IEEEproof}
By Theorem~1  \cite{angelopoulos2023conformal},  {$\max_{k\in[K]}\|\boldsymbol p_{\pi^\star(k)}-\widehat{\boldsymbol p}_k\|_2\leq\widehat q_{1-\alpha}$} holds with probability at least $1-\alpha$. Hence, \eqref{eq:CP_set_construction} gives  {$\boldsymbol p_k\in\mathcal C(\widehat{\boldsymbol p}_{k})$} for all $k\in[K]$.
\end{IEEEproof}}}

\section{Experiment results}
The experiment considers near-field multi-user localization via uplink transmissions with a ULA of $\tilde M=25$ antennas operating at carrier frequency $f=1$\;GHz with half-wavelength antenna spacing. The number $K$ of users is randomly drawn from $K\in\{1,2,3,4\}$. For each user $k\in [K]$, its DoA $\theta_{k0}$ is drawn uniformly as $\theta_{k0}\sim \mathcal{U}({-60}^\circ,60^\circ)$, with a minimum angular separation constraint $\min_{i\neq j}|\theta_{i0}-\theta_{j0}|\ge 10^\circ$, and its distance (in meters) $r_{k0}$ is generated uniformly over a fan-shaped sector, denoted by $r_{k0}^2\sim \mathcal U(5^2,55^2)$. For the NLoS paths $l\in[L_k]$, $k\in[K]$, the number $L_k$ of NLoS paths associated with user $k$ is taken from $\{0,1,2\}$; the DoA $\theta_{kl}$ is generated by $\theta_{kl}\sim \mathcal U(\theta_{k0}-5^\circ,\theta_{k0}+5^\circ)$ with a minimum angular separation no larger than $0.5^\circ$; the corresponding distance $r_{kl}$ is generated by $r_{kl}^2\sim \mathcal U((r_{k0}-10)^2,(r_{k0}+10)^2)$; and the  complex reflection coefficient is generated as $c_{kl}=\rho_{kl}e^{j\phi_{kl}}$ with the phase following $\phi_{kl}\sim \mathcal U(0,2\pi)$ and the amplitude following $\rho_{kl}\sim \mathcal U(0,1)$, respectively. Based on the generated LoS and NLoS parameters, i.e., $\boldsymbol{\Theta}_{\text{LoS}}$ and $\boldsymbol{\Theta}_{\text{NLoS}}$, and given a prescribed transmit $ \mathrm{SNR}=10\log_{10}({\sum_{k=1}^K P_k}/{(K\sigma^2)})$ with equal source powers ($P_1 = P_2 = \cdots = P_K$), the SCM $\boldsymbol{\widehat R}$ is synthesized according to \eqref{eq:SCM} with $T=25$ snapshots. As a result,  we generate a total of $1.25\times 10^5$ data points $(\widehat{\boldsymbol R},  \boldsymbol{\Theta}_{\mathrm{LoS}})$ with $10^5$ for training and $1.25\times 10^4$ each for calibration and test in SCP.

We employ ResNet-based networks~\cite{he2016deep} as the neural architectures for the LoS feature extractor and surrogate distance estimator. Specifically, a two-channel tensor is formed by the real and the imaginary parts of the SCM $\widehat{\mv R}$ as the input, and then is processed by a $3\times3$ convolutional layer followed by $N$ residual stages to extract high-level features. Each residual stage consists of
a shortcut branch with a $1\times1$ convolutional layer and a residual branch
with $L$ residual blocks, where each residual block contains two
``ReLU--$3\times3$ convolution'' layers. The extracted high-level feature is further passed
through a ReLU activation, average pooling, and a fully connected layer. The networks are trained using the Adam optimizer~\cite{kingma:adam} with a mini-batch size of $128$. The test accuracy is evaluated by the MPER, computed by averaging \eqref{eq:PER_score} over all users and all samples in the test dataset.

We consider the following benchmark positioning schemes:  \textbf{{2D MUSIC}}, which directly estimates
the $K$ users' positions via two-dimension angle-distance spectral search;
\textbf{{Two-phase MUSIC}}~\cite{cheng2025comprime}, which applies
spatial smoothing to decorrelate coherent DoAs but does not decorrelate coherent ranges;
\textbf{{NF-SubspaceNet}}~\cite{Gast2026NFsubspace}, which employs an autoencoder to learn a surrogate covariance matrix for subsequent 2D MUSIC-based localization; and \textbf{{DCD-MUSIC}}~\cite{gast2025dcd},
which employs two autoencoders to learn surrogate covariance
matrices for cascaded DoA and range estimation, and further trains the cascaded architecture using positioning errors.

Note that {Two-phase MUSIC}, {NF-SubspaceNet}, and {DCD-MUSIC} are aimed for  estimating all parameters including both $\boldsymbol{\Theta}_{\mathrm{LoS}}$ and $\boldsymbol{\Theta}_{\mathrm{NLoS}}$. To enable a fair comparison with the LoS-positioning objective considered in this paper, we introduce the following two evaluation approaches for these schemes: \textbf{{Indirect LoS}}, by which all path-related parameters are first estimated and then grouped user-wise using the path association strategy in~\cite[Algorithm 1]{xie2017source}; then, for each associated path group, the delay-and-sum algorithm~\cite{widrow2005adaptive} is applied to select the LoS angle-distance pair estimates of the corresponding user; and \textbf{{Direct LoS}}, by which the outputs of the original schemes are replaced by E2E predictions of the multi-user positions.

\begin{figure}[t]
\centering
\includegraphics[width=2.6in]{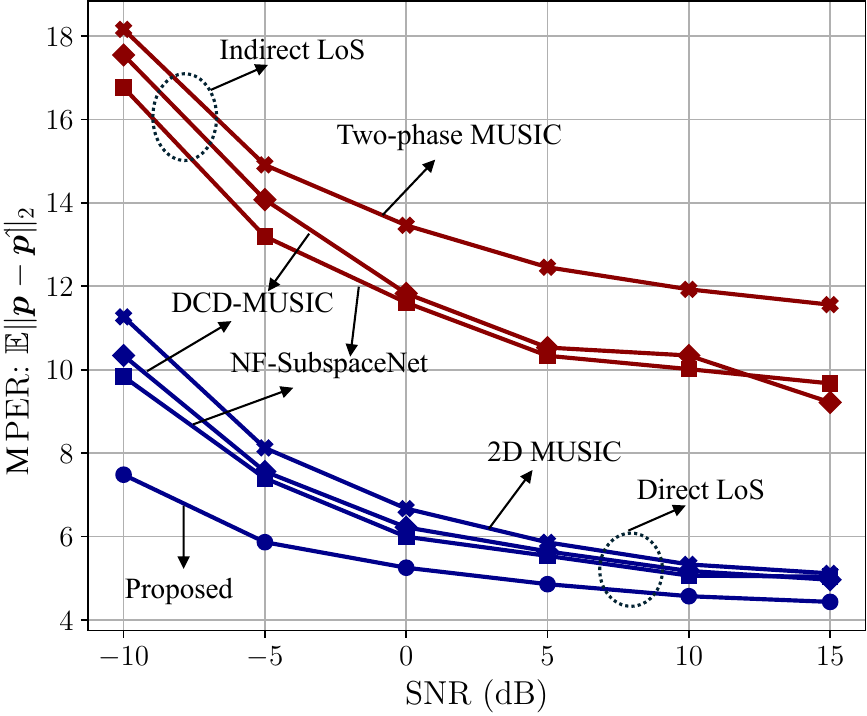}
\caption{The MPER versus  the SNR.}
\vspace{-0.2in}
\label{fig:PER}
\end{figure}

First, we evaluate the point estimation accuracy using the MPER between the 
estimated position $\widehat{\boldsymbol p}$ and the ground truth 
$\boldsymbol p$, defined as 
$\mathbb{E}\|\boldsymbol p-\widehat{\boldsymbol p}\|_2$, where the expectation 
is taken over all samples in the test dataset. Fig.~\ref{fig:PER} shows that the proposed MUSIC-Net outperforms all other methods across the entire SNR 
range. Moreover, the direct-LoS baselines generally outperform their 
indirect-LoS counterparts, indicating that directly targeting the LoS-positioning  objective is more effective than first estimating all paths and then extracting 
the LoS component, due to the accumulated errors introduced in path estimation 
and post-processing.

\begin{figure*}[t]
\centering
   \noindent\hspace*{-0.01\textwidth}%
    \subfigure[$1-\alpha=0.7$]{
        \includegraphics[width=0.32\textwidth]{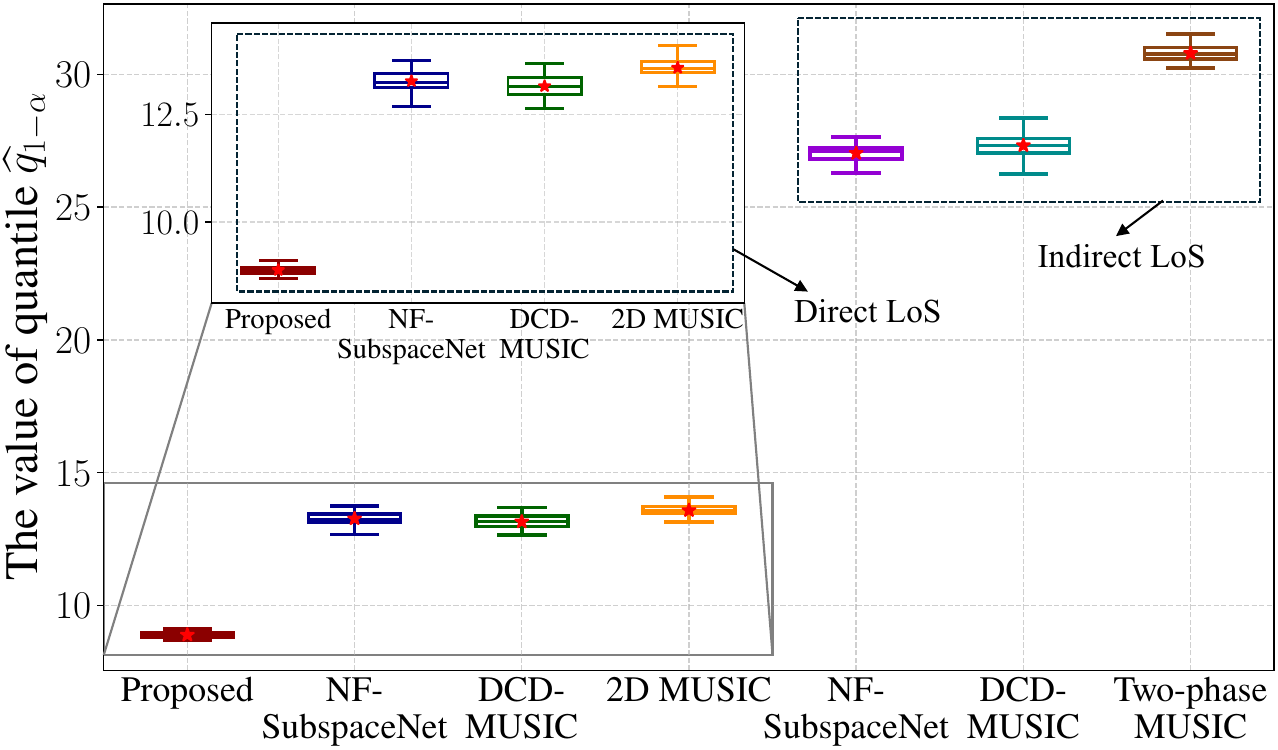}
        \label{fig:box07}
    }
    \hspace{-0.005\textwidth}%
    \subfigure[$1-\alpha=0.8$]{
        \includegraphics[width=0.32\textwidth]{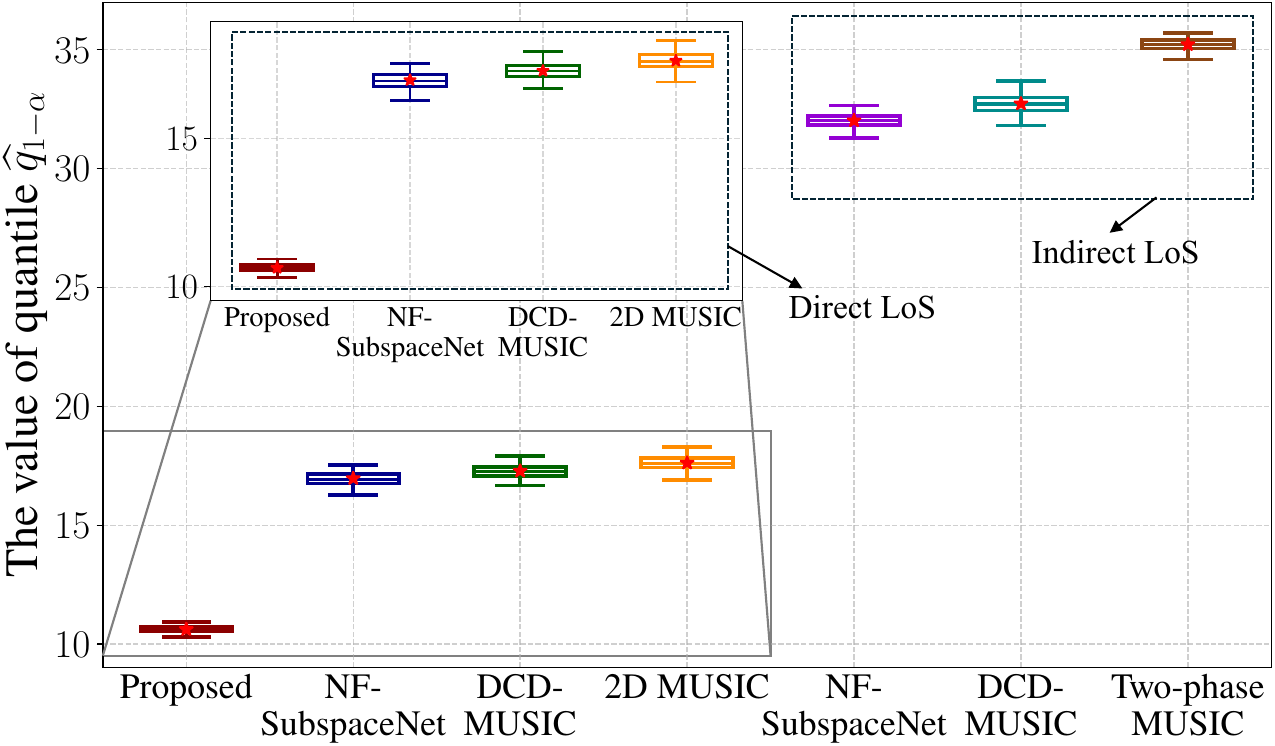}
        \label{fig:box08}
    }
   \hspace{-0.005\textwidth}%
    \subfigure[$1-\alpha=0.9$]{
        \includegraphics[width=0.32\textwidth]{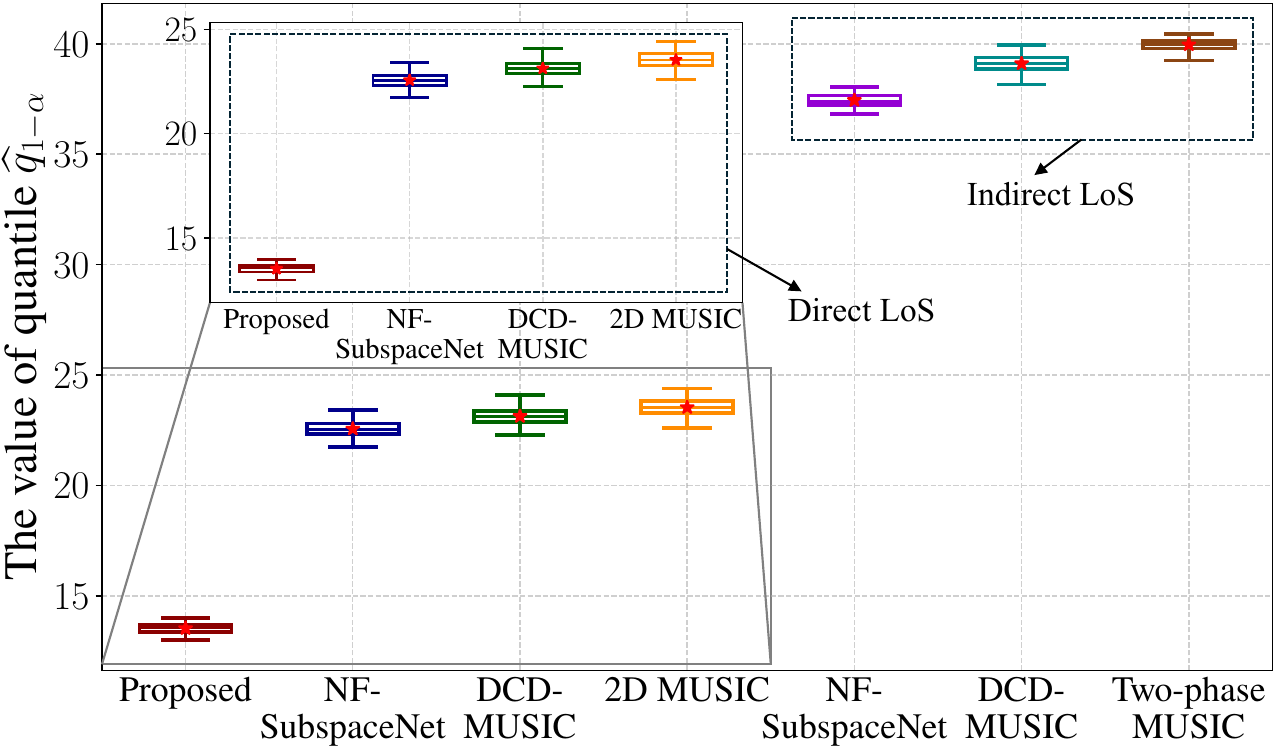}
        \label{fig:box09}
    }
    \vspace{-0.1in}
    \caption{Box plots for the radius $\widehat{q}_{1-\alpha}$ with different target coverage levels ($1-\alpha$),  $\text{SNR}=5$ dB, and \textcolor{red}{$\star$} marks the empirical mean over $100$ Monte Carlo trials.}
    \label{fig:box_all}
\end{figure*}

\begin{figure}[h]
\centering
\includegraphics[width=3in]{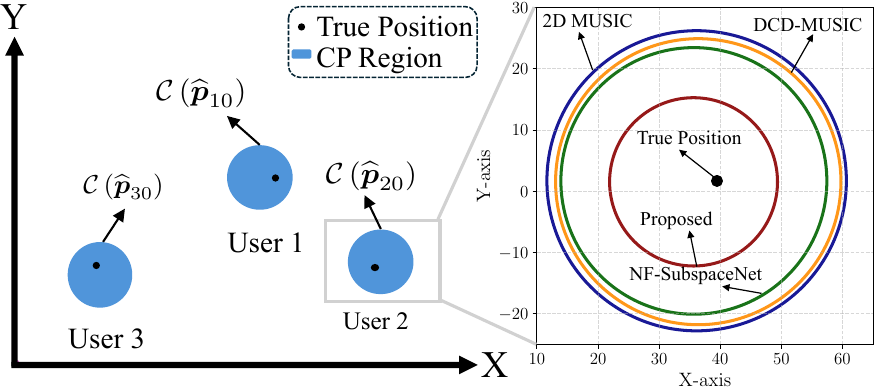}
\vspace{-0.05in}
\caption{{ Visualization of the regions 
$\{\mathcal{C}(\widehat{\boldsymbol p}_k)\}_{k=1}^{K}$ constructed by 
\eqref{eq:CP_set_construction} with $\mathrm{SNR}=5$ dB, $\alpha=0.1$, $K=3$, and  {the region size obtained by different schemes.}}}
\label{fig:location}
\end{figure}

{ Next, Fig.~\ref{fig:location}  {illustrates}  the confidence regions 
$\{\mathcal C(\widehat{\boldsymbol p}_k)\}_{k=1}^{K}$ constructed by 
\eqref{eq:CP_set_construction} for  {all users} in a  {typical} sample. It shows that the proposed MUSIC-Net produces a much  {smaller predicted} region covering the true location $\boldsymbol{p}$  {compared to other benchmark schemes.}}

Finally, we evaluate the efficiency of the confidence regions 
{$\{\mathcal{C}(\widehat{\boldsymbol p}_k)\}_{k=1}^{K}$} under different target coverage levels 
$1-\alpha$ by comparing the value of radius $\widehat{q}_{1-\alpha}$ across 
different benchmarks in Fig.~\ref{fig:box_all}. The results show that the proposed MUSIC-Net 
yields the smallest radius $\widehat{q}_{1-\alpha}$ among all benchmarks, indicating 
that it achieves the tightest  {predicted} regions while satisfying the prescribed coverage level described in \eqref{eq:marginal_joint_coverage}. Moreover, increasing the target coverage level $1-\alpha$ comes at the cost of enlarging the prediction set.

\section{Conclusion}
\label{sec:conclusion}
This paper proposed MUSIC-Net, an E2E DL framework informed by two-stage MUSIC for reliable near-field multi-user positioning in mixed LoS and NLoS multipath environments. The proposed framework integrated a LoS feature extractor and a surrogate distance estimator with two-stage MUSIC and employed subspace-distance-based losses to directly recover user positions of  LoS paths, thereby avoiding redundant parameter estimation for NLoS paths and subsequent path/source association. Furthermore, SCP was employed to augment the point estimates with prediction regions that provided statistical joint coverage guarantees for all users. Numerical results showed that the proposed framework achieved lower MPER than existing benchmarks and yielded tighter SCP-calibrated prediction regions, demonstrating both accurate LoS localization and reliable UQ in coherent multi-path environments.

\balance
\bibliographystyle{IEEEtran}

\bibliography{reference}

\end{document}